\documentstyle[aps,epsf,multicol]{revtex}

\begin{document}
\title{Autosolitons in trapped Bose-Einstein condensates
with two- and three-body inelastic processes}
\author{Victo S. Filho$^1$, F.Kh. Abdullaev$^{1,2}$, A. Gammal$^1$,
Lauro Tomio$^1$ }
\address{$^{1}$Instituto de F\'{\i}sica Te\'orica, Universidade Estadual
Paulista, 01405-900, S\~ao
Paulo, Brazil,\\
${^2}$Physical-Technical Institute, Tashkent, Uzbekistan }
\date{\today}
\maketitle

\begin{abstract}
In this work, we consider the conditions for the existence of autosolitons, 
in trapped Bose-Einstein condensates with attractive atomic interactions.
First, the variational approach is employed to estimate the stationary
solutions for the three-dimensional Gross-Pitaevskii equation with trap
potential, linear atomic feeding from the thermal cloud and two- and
three-body inelastic processes. Next, by using exact numerical calculations, 
we show that the variational approach gives reliable analytical results.
We also discuss the possible observation of autosolitons in experiments 
with $^7$Li.
\newline\newline
PACS numbers: {03.75.Fi, 32.80.Pj, 42.50.Md, 42.81.Dp}
\end{abstract}
\begin{multicols}{2}

\section{Introduction} 
The existence of envelope autosolitons in one-dimensional (1D) case, in
homogeneous nonlinear medium with dissipation and amplification, was
revealed by Pereira and Stenflo\cite{Pereira}. They found the exact
solution for autosolitons with arbitrary growth and damping strengths in
the perturbed nonlinear Schr\"odinger equation (NLSE). Later,
autosolitons were discovered in  nonlinear fiber optics, namely in
fibers with amplifiers and distributed filters (the latter corresponds to
the frequency-dependent damping in the nonlinear Schr\"odinger
equation)~\cite{Agrawal,Abdullaev} and also for waves on the
surface of deep water\cite{Fabrikant}. Correspondingly, in a
two-dimensional (2D) homogeneous medium with amplification and nonlinear
damping, the possibility of existence of a 2D analog of the
Pereira-Stenflo solitons was recently shown by a variational
approach~\cite{Anderson1}. 
Autosolitons in a weakly dispersive nonlinear
media, described by the Korteweg-deVries equation, have been studied
in Refs.~\cite{Abdullaev2,Abdullaev4}.

The autosolitons can be distinguished from  ordinary solitons. The
latter exist in  conservative media and 
are originated from the balance between the nonlinear and dispersive
effects of the wave propagation. The properties of these generated
solitons are defined by the initial conditions (their number, parameters
like amplitudes, widths, etc.)~\cite{Abdullaev2}, with the solutions
characterized by their corresponding properties.
As for to the autosolitons, they can be generated in nonconservative
media when effects of amplification and dissipation are present.
For the existence of autosolitons, one should add to 
the equilibrium condition between nonlinearity and dispersion
the requirement of a balance between amplification,
frequency-dependent damping and nonlinear dissipation.
In distinction from ordinary solitons, the properties of the
autosoliton, as a rule, are fixed by the coefficients of the
perturbed NLSE and by any initial perturbation that is attracted to this
point (attractor in the space of coefficients). Mathematically,
the problem is described by the NLSE with  complex parameters.
If the imaginary parts are large, the equation is equivalent to the
so-called complex Ginzburg-Landau equation.  An interesting limit
is represented by the NLSE with small complex coefficients.

The purpose of this article is to show that the analog of autosoliton is
possible in a trapped  Bose-Einstein condensate (BEC).
Recently  the existence of bright and dark solitons in BEC has
been demonstrated (see theory in \cite{sol1,sol2}).  Dark solitons
have already been observed in BEC with repulsive interaction between atoms
\cite{sol}. For attractive interactions, bright solitons exist in 1D BEC.
It is well known that two- and three-dimensional  condensates with
attractive interaction between atoms and trapping potential are  unstable
if the number of atoms exceeds the critical value ($N_c$). Below this
value a stable ground state can exist, corresponding to
solitary solutions \cite{Gammal1,Kivshar}. When the number of
atoms exceeds $N_c$ the collapse occurs. At large densities the
inelastic scattering processes involving two and three atoms come
into play, leading to the effective nonlinear damping of the
condensate. The process of feeding atoms from the thermal cloud
can be modeled as a linear amplification described in Ref.~\cite{Kagan},
where the relevance of the three-body inelastic processes was 
discussed.
The statistical data obtained from experiments with $^{7}$Li 
supports the growing and collapse picture \cite{Hulet,Stoof}.

Numerical simulations of the 3D Gross-Pitaevskii (GP) equation are
performed in Refs. \cite{Kagan,Gammal}. 
The present work shows that
periodic oscillations occur in the condensate and, for particular
values of atomic feeding and three-body dissipation parameters,
stable states of the cloud can exist. Thus, we can expect the 
occurrence of analog autosolitons in 2D and 3D BECs.
The problem is described by the complex Ginzburg-Landau equation
with trapping potential in the NLSE limit with small
nonconservative perturbations. This equation is nonintegrable and,
therefore, analytical solutions can only be obtained by considering 
approximate methods like the variational approach 
\cite{Kaup}.
In the following, we first use the time-dependent variational approach 
to obtain the solutions.  As showed previously, the
time-dependent variational approach is quite effective to study
the dynamics of 3D BEC in trapping potential with conservative
perturbations \cite{Garcia}. These autosoliton solutions can be
considered as the nonlinear modes of such system like solitons for
the integrable NLSE \cite{Akhmediev}. Exact numerical simulations are 
also performed in the present work, which confirms that, in the
present case, the variational approach is a convenient and reliable
approximation.

\section{The model and the variational approach}
Let us describe 
the dynamics of a trapped Bose-Einstein condensate in the framework
of the Gross-Pitaevskii equation.
The space-time dynamics of the condensate wavefunction can be 
analyzed by the GP equation in the mean field approximation:
\begin{eqnarray}\label{GP} 
{\rm i}u_t + \Delta u - (\Omega^2 r^2) u + \lambda_2 |u|^2 u +
\lambda_3 |u|^4 u \nonumber \\ = 
{\rm i}\left(\gamma-\mu |u|^2-\xi|u|^4\right)u = R(u,u^* ), 
\end{eqnarray} 
where we use a standard simplified notation for the time derivative
through a lower index $t$. $\Omega^2 r^2 $ is the trap harmonic 
potential; $\lambda_2$ and $\lambda_3 $ are, respectively, the two and
three body interaction parameters, where $\lambda_2 (= a_s)$ is given 
by the $s$- wave atomic scattering length. $\gamma$, $\mu$, and $\xi$ are
positive defined coefficients related, respectively, to feeding, dipolar
relaxation and three-body  inelastic recombination parameters. We drop the
explicit time and radial dependence of the functions, unless they are
necessary for clarity.

In the present variational approach, for $u\equiv u(r,t)$, we use the
Gaussian trial function\cite{Anderson2}
\begin{equation}\label{tr}
u = A(t) \exp\left( -\frac{r^2}{2a^2(t)} + 
{\rm i}\frac{b(t)r^2}{2}+{\rm i}\phi(t)\right) ,
\end{equation}
where $A$, $a$, $b$, and $\phi$ are, respectively, the amplitude, width,
chirp and linear phase. We did not include the center-of-mass
coordinate  into  the ansatz, because the dissipative and 
amplifying terms have no  influence on it. 

The variational approach is applied to the averaged Lagrangian of
the conservative system
\begin{equation}\label{av}
L = \int {\cal L}(r,t)d^3 r,
\end{equation}
where the Lagrangian density, ${\cal L}\equiv {\cal L}(r,t)$, is given by
\begin{equation}\label{L}
{\cal L} = \frac{\rm i}{2}(u_t u^* - u_t^* u) -|\nabla u|^2 +
\frac{\lambda_2}{2}|u|^4 + \frac{\lambda_3}{3}|u|^6 - \Omega^2 r^2 |u|^2 .
\end{equation}
Substituting the trial function (\ref{tr}) into Eqs. (\ref{L}) and
(\ref{av}), we find the averaged Lagrangian in terms of the condensate
wavefunction  parameters 
\begin{eqnarray}\label{lav}
L = -\frac{\pi\sqrt\pi}{4}A^2a^3
&&\left[3a^2 b_t + 4\phi_t + \frac{6}{a^2}(1+ a^4b^2)\right.
\nonumber\\ 
&&-\left.\frac{\lambda_2}{\sqrt{2}}A^2 -
\frac{4\lambda_3}{9\sqrt 3}A^4 + 6\Omega^2 a^2\right].
\end{eqnarray}

We formally add ${\cal L}_R$ to Eq.(\ref{L}), with the property that
$\delta {\cal L}_R/\delta u^*=-R(u,u^*)$, 
where $R$ is the right hand side of Eq.(\ref{GP}). 
By applying the Euler-Lagrange equations to 
${\cal L}'\equiv {\cal L} + {\cal L}_R$, with respect to 
$u^*$, we obtain 
\begin{eqnarray}\label{Lprime}
\left[\frac{\partial {\cal L^\prime}}{\partial u^*} -
\frac{d}{dt}\frac{\partial{\cal L^\prime}}{\partial u^*_t}\right] 
= \left[\frac{\partial {\cal L}}{\partial u^*} -
\frac{d}{dt}\frac{\partial{\cal L}}{\partial u^*_t}\right] 
- R(u,u^*) = 0 ,
\end{eqnarray}
that leads to Eq.~(\ref{GP}) (The conjugate equation is obtained
in a similar way.).

The corresponding variational principle is given by
\begin{equation}
\delta\int_0^t L'dt = \delta\int_0^t (L + L_R) dt = 0 ,
\end{equation}
where, as in Eq.(\ref{av}), $L_R = \int d^3 r {\cal L}_R .$ 
Taking into account that for a small shift $\delta\eta$ of some variational 
parameter $\eta$, we have
\begin{equation}
f(\eta + \delta\eta) = f(\eta) + \delta\eta
\frac{\partial f}{\partial\eta},
\end{equation}
where $f\equiv f(u,u^\ast) = L$ or $L_R$, 
we obtain a system of equations for the variational parameters 
$\eta_i$~\cite{Anderson2,Maimistov}:

\begin{equation}\label{nc}
\frac{\partial L}{\partial \eta_i} -
\frac{d}{dt}\frac{\partial L}{\partial\eta_{it}} = \int
d^3 r \left[R\frac{\partial u^*}{\partial\eta_i} + R^*\frac{\partial
u}{\partial\eta_i}\right],
\end{equation}
where Eq.~(\ref{Lprime}) and its conjugate were used.

The substitution of  Eqs.(\ref{tr}) and (\ref{lav}) into Eq.(\ref{nc}),
yields the following system of ODE's :

\begin{eqnarray}
\frac{d(A^2 a^3)}{dt} &=& 2\gamma A^2 a^3  - \frac{\mu}{\sqrt{2}}A^4
a^3 - \frac{2}{3\sqrt 3}\xi A^6 a^3, \nonumber\\
\frac{d(A^2 a^5 )}{dt}&=& 4 A^2 a^5 b + 2\gamma A^2 a^5 -
\frac{\mu}{2\sqrt{2}}A^4 a^5 - \frac{2}{9\sqrt{3}}\xi A^6 a^5,
\nonumber\\ 
\frac{db}{dt} &=& \frac{2}{a^4} - 2b^2
-2\Omega^{2} - \frac{\lambda_2A^2}{2\sqrt{2}a^2}
- \frac{4\lambda_3A^4}{9\sqrt{3} a^2}, \nonumber\\
\frac{d\phi}{dt} &=& -\frac{3}{a^2}+\frac{7}{8\sqrt{2}}\lambda_2
A^2+\frac{2}{3\sqrt{3}} \lambda_3 A^4 .
\label{sys}
\end{eqnarray}

The Eq. (\ref{sys}a) can also be obtained from the modified form of the
conservation law for the number of atoms $N$, where $N$ is given by
\begin{equation}
N = \int |u|^2 d^3r \; .\label{number}
\end{equation}
The other equations of the system (\ref{sys}) can be obtained  
by using higher moments, as shown in the appendix.
It is useful to rewrite the system using the notation
$x = a^2$, $y  = A^2$:
\begin{eqnarray}\label{sys1}
x_t &=& 4xb +\frac{\mu}{2\sqrt{2}}xy + \frac{4}{9\sqrt{3}}\xi y^2 x,
\nonumber \\ y_t &=& -6yb + 2\gamma y - \frac{7\mu}{4\sqrt{2}}y^2
- \frac{4}{3\sqrt{3}}\xi y^3, \nonumber\\
b_t &=& \frac{2}{x^2} - 2b^2 - 2\Omega^2 - \frac{\lambda_2y}{2\sqrt{2}x}
-\frac{4\lambda_3 y^2}{9\sqrt{3}x}.
\end{eqnarray}
This ODE system is the main result of this section.

\section{Analysis of the fixed points}
The autosoliton solution corresponds to the fixed points of the
system.
Recall the main properties of the autowave (dissipative) soliton.
This solution has the form like for standard soliton, 
defined by the balance between amplification and
nonlinear dissipative terms. 
As showed in the analysis of the one dimensional case (the Pereira-Stenflo
soliton), 
the solution is fixed by the parameters related to the amplification and
dissipation with chirped phase. It does not depend on the initial conditions.

We restrict our analysis to $\lambda_3=0$. From the system given by 
Eq. (\ref{sys1}), we can obtain the fixed points. In the next we
distinguish three cases:
\newline\newline
\noindent{\bf 1. Case $\mu= 0$, $\xi \neq 0$.} 

In this case,
\begin{equation}\label{fp}
y_{s1} = \sqrt{\frac{3\sqrt{3}\gamma}{\xi}},\;\;\; b_{s1} =
-\frac{1}{3}\gamma,
\end{equation}
and the width is
\begin{equation}\label{fp1}
x_{s1} = -\frac{p_1}{2} + \sqrt{\frac{p_1^2}{4}+ k_1},
\end{equation}
where 
\begin{equation}
p_1 = \frac{\lambda_2 y_{s1}}{4\sqrt{2}(\Omega^2 + b_{s1}^2 )},
\;\;\;\; k_1 = \frac{1}{\Omega^2 + b_{s1}^2}.
\end{equation}

Let us consider that $p_1>> k_1$. Then the
solution is $$x_{s1} \approx \frac{k_1}{p_1} \approx
\frac{4\sqrt{2}\Omega^4}{\lambda_2 y_{s1}}.$$
\newline
\noindent{\bf 2. Case $\xi = 0$, $\mu \neq 0$.} 

The fixed points are
\begin{equation}\label{fp2}
y_{s2} = \frac{2\sqrt{2}\gamma}{\mu}, \ b_{s2} =
-\frac{\gamma}{4}.
\end{equation}
and the width is defined by the same expression as before
Eq.(\ref{fp1})with $y_{s2}, b_{s2}$.
\newline\newline
\noindent{\bf 3. Case $\mu\neq 0$, $\xi \neq 0$.}

The fixed points are:
\begin{eqnarray}
y_{s3} &=& -\frac{3\sqrt{3}\mu}{4\sqrt{2} \xi } +
\sqrt{\frac{27 \mu^2}{32 \xi^2} +
\frac{3\sqrt{3}\gamma}{\xi}},\nonumber\\
b_{s3} &=& \frac{\mu}{24\sqrt{2}}y_{s3} - \frac{\gamma}{3},\nonumber\\
x_{s3}&=& -\frac{p_3}{2} + \sqrt{\frac{p_3^2}{2} + k_3},
\end{eqnarray}
where 
\begin{equation}
p_3 = \frac{\lambda_2 y_{s3}}{4\sqrt{2}(\Omega^2 + b_{s3}^2 )},
\;\;\;\;\; 
k_3 = \frac{1}{\Omega^2 + b_{s3}^2}.
\end{equation}
When $\mu = 0$ ($\xi \neq 0$), we
recover Eqs.(\ref{fp}) and (\ref{fp1}). Alternatively, when
$\xi\rightarrow 0$ ($\mu \neq 0$) we recover Eq.(\ref{fp2}).

We now investigate the stability of the fixed points, in cases 1 and 2, 
by using the linear stability analysis:
\newline\newline
\noindent{\bf 1. Case $\mu = 0$, $\xi \neq 0$.}

Let present the solutions of the system as 
$x=x_{s1} + x_1$, $y = y_{s1} + y_1$, $b = b_{s1} + b_1$. 
The linearized system for corrections is
\begin{eqnarray}\label{corr1}
x_{1t} &=& \frac{8}{9\sqrt{3}}\xi y_{s1} x_{s1} y_1 + 4x_{s1} b_1 
\equiv -c_2 y_1 + c_3 b_1 ,\nonumber\\
y_{1t} &=& - 8\gamma y_1 - 6y_{s1} b_1\equiv -d_2 y_1 - d_3 b_1 ,
\nonumber\\
b_{1t} &=& \frac{4}{x_{s1}^2}\left(\frac{\lambda_2 y_{s1}}{8\sqrt{2}} -
\frac{1}{x_s}\right)x_1
- \frac{\lambda_2}{2\sqrt{2}x_{s1}}y_1 - 4b_{s1}
b_1 \nonumber \\ &\equiv& a_1 x_1-a_2 y_1 - a_3 b_1 .
\end{eqnarray}
With the solutions $x_1$, $y_1$ and $b_1$ having the same
exponential behavior in time, given by $\sim e^{qt}$, we obtain
the characteristic equation
\begin{equation}\label{cubic1}
q^3 + \alpha_1 q^2 - \alpha_2 q - \alpha_3 = 0,
\end{equation}
where
\begin{eqnarray}
\alpha_1 &\equiv& (d_2 + a_3) = \frac{20}{3}\gamma, \nonumber\\ 
\alpha_2 &\equiv& (a_1 c_3 + a_2 d_3 - d_2 a_3)\nonumber\\ 
         &=& \frac{32}{3}\gamma^2 +
\frac{16}{x_{s1}^2}\left(\frac{\lambda_2 y_{s1} x_{s1}}{8\sqrt{2}} -
1\right) + \frac{3\lambda_2 y_{s1}}{\sqrt{2}x_{s1}},
\nonumber \\ 
\alpha_3 &\equiv& a_1(d_3 c_2 + d_2 c_3 ) =
\frac{64\gamma}{x_{s1}^2}\left(\frac{\lambda_2 y_{s1} x_{s1}}{8 \sqrt{2}}
- 1\right).
\end{eqnarray}
The roots with Re$(q) > 0$ correspond to the unstable solutions.

 \begin{figure}
 \setlength{\epsfxsize}{1.0\hsize}
\centerline{\epsfbox{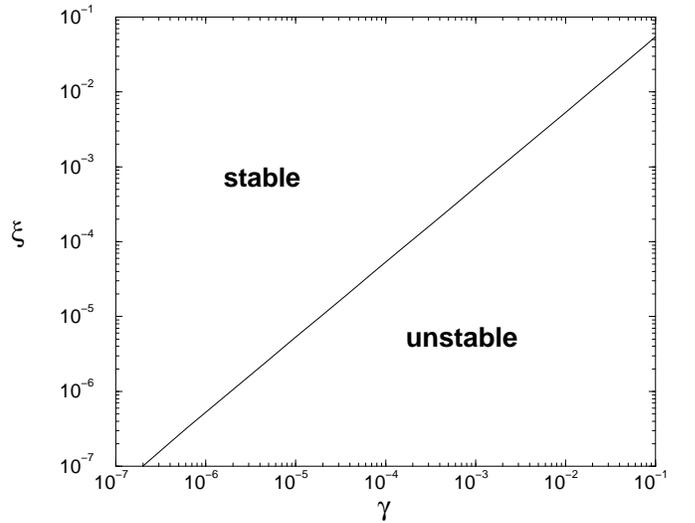}}
 \caption[dummy0]{
Gaussian variational analysis of stability of the fixed points for the
Gross-Pitaevskii equation including feeding ($\gamma$) and three-body
losses ($\xi$).
   } \end{figure}
Without loss of generality, in our dimensionless NLSE we can
scale the parameters as $\lambda_2=1$ and $\Omega=1$
\cite{Kagan}. \ The diagram of stability, according to the solutions of
Eq.~(\ref{cubic1}), is presented in Fig. 1. 
The diagram clearly shows that, when $\gamma > 1.84 \xi$, the system is
unstable. If $\gamma >> \xi$, the system enters the collapsing process
that has been shown in Ref.~\cite{Gammal} to be chaotic. If
$\gamma$ is decreased (or $\xi$ increased) the system will eventually
achieve a stable region where the formation of autosoliton is possible.
\newline\newline

\noindent{\bf 2. Case $\xi = 0, \mu \neq 0$.}

Analogously, as in case 1, we present the solutions of the system as
$x=x_{s2} + x_2$, $y= y_{s2} + y_2$, and $b = b_{s2} + b_2$.
The linearized system for corrections is given by
\begin{eqnarray}\label{corr2}
x_{2t} &=& \frac{\mu x_{s2}}{2\sqrt{2}}y_2 + 4x_{s2} b_2,\nonumber\\
y_{2t} &=& -\frac{7}{2}\gamma y_2 - \frac{12\sqrt{2}\gamma}{\mu}b_2 ,
\nonumber\\
b_{2t} &=& \frac{4}{x_{s2}^3}\left(\frac{\lambda_2 \gamma}{4\mu}x_{s2}
- 1\right)x_2
-\frac{\lambda_2}{2\sqrt{2}x_{s2}}y_2 - 4b_{s2} b_2 .
\end{eqnarray}

The system has the same form as in Eq. (\ref{corr1}). 
Correspondingly, the characteristic equation is given by
\begin{equation}\label{cubic2}
q^3 + \beta_1 q^2 - \beta_2 q - \beta_3 = 0,
\end{equation}
where
\begin{eqnarray}
\beta_1 &=& \frac{5}{2}\gamma, \nonumber\\
\beta_2 &=& \frac{7}{2}\gamma^2 +
\frac{16}{x_{s2}^2}\left(\frac{\lambda_2 \gamma x_{s2}}{4 \mu } - 1
\right)+
\frac{6\lambda_2 \gamma}{x_{s2} \mu} ,  \nonumber \\ 
\beta_3 &=& \frac{32\gamma}{x_{s2}^2}\left(\frac{\lambda_2 \gamma
x_{s2}}{4 \mu } - 1\right).
\end{eqnarray}
 \begin{figure}
 \setlength{\epsfxsize}{1.0\hsize}
\centerline{\epsfbox{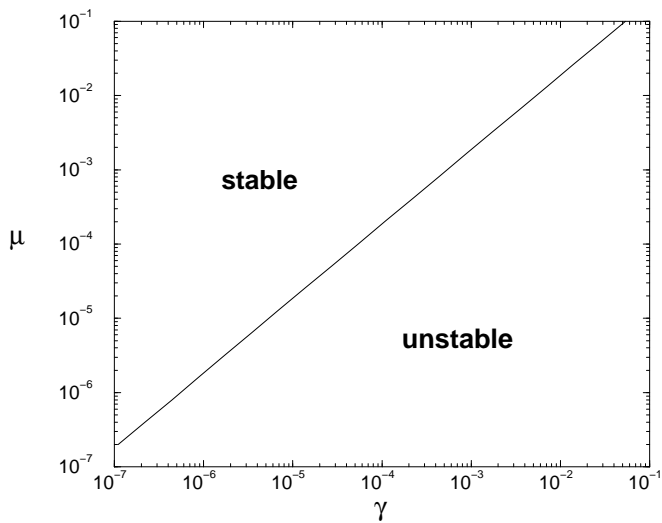}}
 \caption[dummy0]{
Gaussian variational analysis of stability of the fixed points for the
Gross-Pitaevskii equation including feeding ($\gamma$) and two-body
losses($\mu$).
   } \end{figure}
With the same scaling used in case 1 ($\lambda_2=1$ and $\Omega=1$), and
with the above solutions of Eq.(\ref{cubic2}), we obtain the diagram of
stability, shown in Fig. 2. 
The diagram clearly shows that, when $\gamma > 0.53 \mu$, the system is
unstable. In analogy with the previous case, if $\gamma$ is decreased (or
$\mu$ increased), the system will eventually achieve a stable region
where the formation of autosoliton is possible.

\section{Numerical simulations and discussions} 
We did a series of time-dependent simulations of the system within the
Gaussian variational approach, using Eq.(\ref{sys1}), 
and also by performing exact numerical calculations with Eq.(\ref{GP}).
In our numerical calculations, we have used the finite-difference
Crank-Nicolson algorithm. The exact initial wave functions were used 
following the prescription given in \cite{Gammal2}. In the
next, we present simulations in a range of parameters that lead
to long-time stable autosolitons. We obtain results with autosolitonic
solutions for a class of parameters that are near the realistic ones,
as indicated by $^7$Li experiments in BEC. 

 \begin{figure}
 \setlength{\epsfxsize}{1.0\hsize}
\centerline{\epsfbox{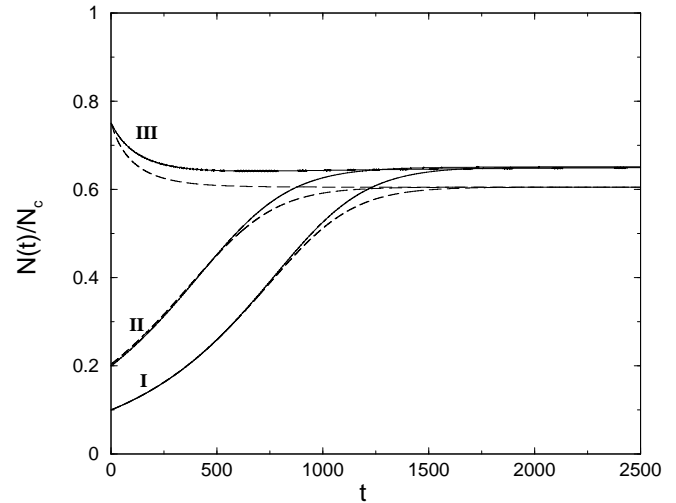}}
 \caption[dummy0]{ Evolution of the number of atoms $N$ in the
Gross-Pitaevskii equation with feeding parameter $\gamma=10^{-3}$ and
three-body dissipation parameter $\xi=10^{-3}$ ($\mu=0$). 
The results are represented in dashed lines for the variational approach,
and in solid lines for the exact numerical calculations.
Cases I, II and III correspond, respectively, to the initial conditions
$N(t=0)/N_c$=0.1, 0.2 and 0.75, where $N_c$ is the maximum critical
number for stability. $t$ is given in units of the inverse of the 
trap frequency $\Omega$.
   } \end{figure}

In Fig. 3, for $\gamma=10^{-3}$ and $\xi=10^{-3}$,
we show the time evolution of the number of atoms, in terms of the
maximum critical number for stability, $N_c$.  The formation of the
autosoliton is demonstrated either by Gaussian variational
approach or by exact numerical calculations. 
There is a remarkable agreement between both approaches. Note that the
number of atoms does not depend on the initial conditions, but is related
to the equilibration of the feeding and dissipation. The variational
approach results give a little higher number of atoms than the exact
calculations. 

 \begin{figure}
 \setlength{\epsfxsize}{1.0\hsize}
\centerline{\epsfbox{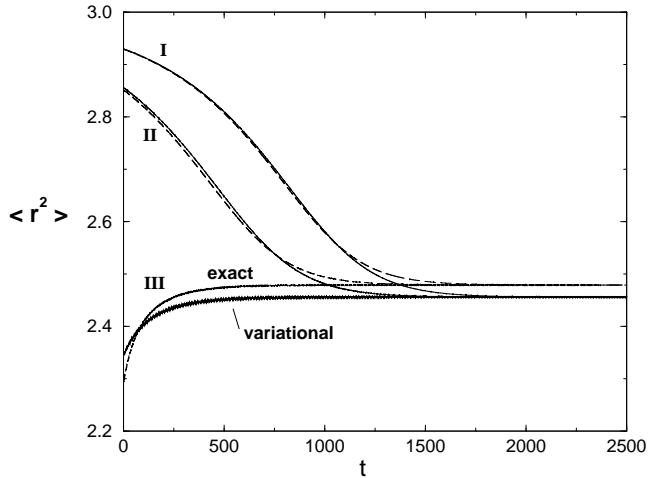}}
 \caption[dummy0]{ Evolution of the mean square radius, Eq.~(\ref{raio}),
in the Gross Pitaevskii equation. The same parameters and conventions
given in Fig. 3 were used. 
} \end{figure}

Corresponding to the results in Fig. 3, we show in Fig. 4
the results for the time evolution of the mean square radius, where 
\begin{equation} 
\langle r^2\rangle =\int r^2 |u|^2d^3r .\label{raio}
\end{equation}
In this case, the variational approach results are a bit lower
than the ones obtained by exact calculations.

In analogy with the case that $\mu=0$, represented in Figs. 3 and 4, we
also present results for the case that the three-body dissipation
parameter ($\xi$) is zero. The results obtained for 
the time evolution of the number of atoms and the mean-square-radius
are, respectively, shown in Figs. 5 and 6, for $\gamma=5\times 10^{-5}$
and $\mu=10^{-4}$.  The formation of the
autosoliton is also demonstrated either by Gaussian variational
approach or by exact numerical calculations. 
The remarkable agreement between both approaches, already observed in 
case that $\mu=0$ (Figs. 3 and 4), also apply to this case that
$\xi=0$, with the number of atoms not depending on the initial
conditions.

In  $^7$Li experiment the feeding parameter can be indirectly inferred
from measurements done by the Rice Group \cite{Hulet} and 
will correspond to an average rate of about 600 atoms/s \cite{Hulet2}.
The two- and three-body losses were also measured \cite{Gerton} and
estimated~\cite{Moerdijk}, giving atom loss rates of about
$2\times 10^{-14}$cm$^3$/s and $\sim 10^{-28}$cm$^6$/s. These 
rates were measured for non-condensed atoms. For condensed atoms they
must be divided by factors of 2! and 3!, respectively \cite{Svistunov}.
With a scaled equation, such that $\lambda_2=1$ and $\Omega=1$, 
we have for condensed atoms $\gamma \sim 10^{-3}$, $\mu \sim 10^{-4}$
and $\xi \sim 10^{-6}$.  Considering this magnitudes, the autosoliton
can possibly be observed experimentally in $^7$Li, either by results with
decreasing $\gamma$ due to losses, or by increasing the
dissipation rate due to other mechanism as the Feshbach
resonances\cite{Cornish}. In case of diminishing $\gamma$, the
autosoliton formation is more likely determined by dipolar
relaxation rather than three-body recombination losses.

 \begin{figure}
 \setlength{\epsfxsize}{1.0\hsize}
\centerline{\epsfbox{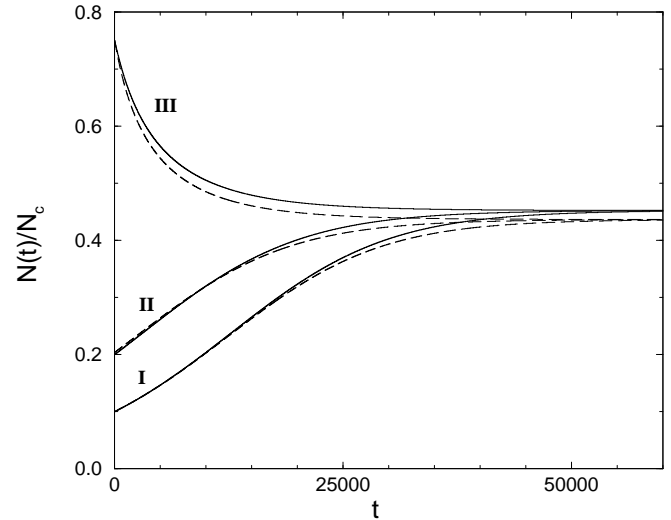}}
 \caption[dummy0]{ Evolution of the number of atoms in
the Gross-Pitaevskii equation with feeding $\gamma=5\times 10^{-5}$ and
two-body dissipation parameter $\mu=10^{-4}$ ($\xi=0$). 
The initial conditions and conventions are the same as in Fig. 3.}
\end{figure}

 \begin{figure}
 \setlength{\epsfxsize}{1.0\hsize}
\centerline{\epsfbox{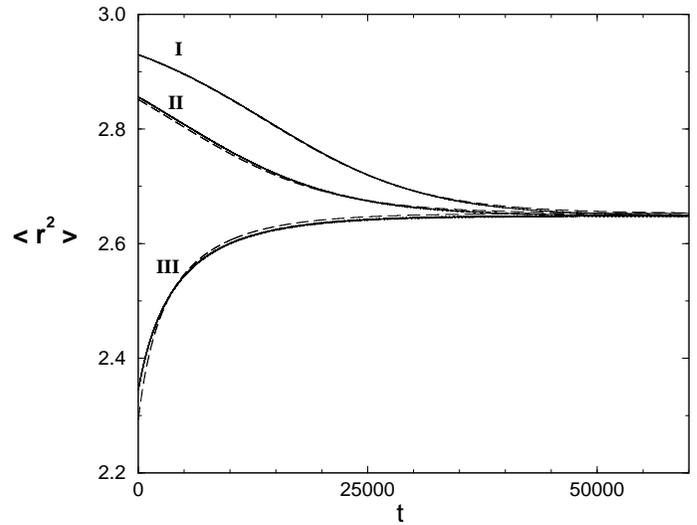}}
 \caption[dummy0]{
Evolution of the mean square radius in the Gross Pitaevskii equation.
The parameters and conventions are the same as in Fig. 5.} 
\end{figure}

\section{Conclusion} 
In this work we study the possibility of existence of autosolitons in
trapped 3D BEC in the presence of two- and three-body inelastic processes:
dipolar relaxation and three-body recombination. Using the time-dependent
variational approach for nonconservative 3D Gross-Pitaevskii equation, we
derive expressions for the parameters of autosoliton and check their
stability. The results obtained by using the present time-dependent
Gaussian variational approach, in the NLSE with atomic feeding and
nonlinear dissipative terms, show a remarkable agreement with exact
numerical calculations, when the parameters were such that stabilization was
achieved. We should note that, stabilization can also be numerically
observed in the model given in Ref.\cite{Kagan}, where $\gamma$ is
decreased in time, but their parameters are far from the realistic ones,
and dipolar relaxation was neglected. Such stabilization verified in 
Ref.\cite{Kagan} has not been recognized as a manifestation of
autosoliton, a characterization that we are pointing out in the
present work.

We have shown that the transition from unstable (collapsing) to stable
point (autosoliton) solely depends on the magnitude of the parameters.
Also the present work include non negligible two-body dissipative effects
that model the dipolar relaxation losses, and that can be associated with
values measured in ultracold $^7$Li\cite{Gerton}. In case of decreasing
$\gamma$ due to collapsing and losses, the autosoliton is more likely to
be formed first due to dipolar relaxation rather than by three-body
recombination processes. These results can be relevant in current
experiments with attractive scattering length and possibly display new
phenomenon of Pereira-Stenflo type autosoliton formation in Bose
condensates. We believe that such phenomenon is already occurring in the
long time
behavior in the actual experiments with $^7$Li \cite{Hulet} ($\Omega \sim
2\pi \times$ 140Hz), since for longer times ($\sim$60s) the maximum number
of atoms ($N_c\sim$ 1300 atoms) is considerably reduced, as expected in
our simulations. We hope that experiments with direct observation of the
evolution of the condensate can clarify this matter.

\section*{Appendix}
The first equation of the system (\ref{sys}) can be obtained by 
calculating the rate of change of the number of atoms as
\begin{equation}\label{dndt}
\frac{dN}{dt} =\frac{d}{dt} \int d^3 r |u|^2 
= \int (u^*_tu+u^*u_t) d^3 r.
\end{equation}
Taking $u_t$ from Eq.~(\ref{GP}) and substituting in (\ref{dndt}) we
obtain the modified form of the conservation law for the number of 
atoms~\cite{Nnew}
\begin{equation}
\frac{dN}{dt} = 2\gamma N - 2\mu\int |u|^4 d^3 r - 2\xi \int |u|^6 d^3r.
\end{equation}
Substituting into this equation the Gaussian trial function (\ref{tr}) we
obtain the first equation of the system given in Eq. (\ref{sys}).

We have derived the Eqs. (\ref{sys}b) and (\ref{sys}c) by the  moments
method. Eq. (\ref{sys}b)  can be derived by calculating
\begin{equation}
\frac{d \left< r^2 \right>}{dt} =\int (u^{\ast}_{t} r^2u+u^*r^2u_t) d^3r.
\end{equation}
Substituting the $u_t$ from expression (\ref{GP})
and applying the commutation rules, we get \cite{Wadati}
\begin{eqnarray}
\label{dpr2dt}
\frac{d \left< r^2 \right>}{dt} = 4\text {Im}\int u^{\ast}
\vec{r} \cdot{} (\nabla u) d^3r
+ 2\gamma\int |u|^2 r^2d^3r
\nonumber \\
- 2\mu \int |u|^4 r^2d^3r-2\xi \int |u|^6 r^2 d^3r .
\end{eqnarray}
The substitution of the Gaussian ansatz (Eq. \ref{tr}) in both
sides of this expression results directly in Eq.(\ref{sys}b).

Equation (\ref{sys}c) can be derived analogously by showing that
\begin{eqnarray}
\label{dp2dt}
\frac{d \left< p^2 \right>}{dt}=
-4\Omega^2\text{Im}\int u^{\ast} \vec{r} \cdot (\nabla u)d^3r \nonumber \\
+2\text{Im}\int (\nabla u^{\ast}) \cdot \nabla (V_1u) d^3r
\nonumber \\
+2\text{Re}\int (\nabla u^{\ast}) \cdot \nabla (V_2u) d^3r
\end{eqnarray}
where
$V_1=-\lambda_2|u|^2-\lambda_3|u|^4$ and
$V_2=\gamma-\mu|u|^2-\xi|u|^4$.
Substituting the Gaussian ansatz (Eq. \ref{tr})  in both sides of Eq.
(\ref{dp2dt})  and using results (\ref{sys}a) and (\ref{sys}b) there is
overall cancellation of the feeding and dissipative terms and we finally
get Eq. (\ref{sys}c).

\section*{Acknowledgements} 
AG is grateful to Prof. R.G. Hulet for useful discussions. 
F.Kh. Abdullaev is grateful to Instituto de F\'\i sica Te\'orica
for the hospitality, and also to partial support of US CRDF (Award
ZM2-2095).
We thank Funda\c{c}\~ao de Amparo \`a Pesquisa do Estado de
S\~ao Paulo (FAPESP) for partial financial support.  
LT also thanks partial support from Conselho Nacional de Desenvolvimento
Cient\'{\i}fico e Tecnol\'ogico (CNPq). 
VSF thanks support received from Coordena\c c\~ao de 
Aperfei\c coamento de Pessoal de N\'\i vel Superior (CAPES).


\end{multicols}
\end{document}